\documentclass[aps,nofootinbib,superscriptaddress, twocolumn, nopacs]{revtex4}
\usepackage[utf8]{inputenc}
\usepackage{graphics}
\usepackage{amssymb}
\usepackage{bm}
\usepackage{braket}
\usepackage{booktabs}
\usepackage{subfloat}
\usepackage{float}

\usepackage{amsmath}    
\usepackage{graphicx}   
\usepackage{verbatim}   
\usepackage{color}      
\usepackage{subfigure}  
\usepackage{hyperref}   
\usepackage{algorithm,algorithmic}

\newcommand{\be}{\begin{equation}}
\newcommand{\ee}{\end{equation}}
\newcommand{\ba}{\begin{eqnarray}}
\newcommand{\ea}{\end{eqnarray}}

\begin{document}

\title{Improving nonstoquastic quantum annealing with spin-reversal transformations}
\author{Eleni Marina Lykiardopoulou}
\affiliation{D-Wave Systems Inc., 3033 Beta Avenue, Burnaby, BC V5G 4M9, Canada}
\affiliation{Department of Physics and Astronomy, University of British Columbia,
Vancouver BC V6T 1Z1, Canada}
\author{Alex Zucca}
\affiliation{D-Wave Systems Inc., 3033 Beta Avenue, Burnaby, BC V5G 4M9, Canada}
\affiliation{Department of Physics, Simon Fraser University, Burnaby BC V5A 1S6, Canada}
\author{Sam A. Scivier}
\affiliation{D-Wave Systems Inc., 3033 Beta Avenue, Burnaby, BC V5G 4M9, Canada}
\affiliation{School of Physics and Astronomy, University of Birmingham, Birmingham B15 2TT, United Kingdom}
\author{Mohammad H. Amin}
\affiliation{D-Wave Systems Inc., 3033 Beta Avenue, Burnaby, BC V5G 4M9, Canada}
\affiliation{Department of Physics, Simon Fraser University, Burnaby BC V5A 1S6, Canada}
\date{\today}

\begin{abstract}
Nonstoquastic Hamiltonians are hard to simulate due to the sign problem in quantum Monte Carlo simulation. It is however unclear whether nonstoquasticity can lead to advantage in quantum annealing. Here we show that YY-interactions between the qubits makes the adiabatic path during quantum annealing, and therefore the performance, dependent on spin-reversal transformation. With the right choice of spin-reversal transformation, a nonstoquastic Hamiltonian with YY-interaction can outperform stoquastic Hamiltonians with similar parameters. We introduce an optimization protocol to determine the optimal transformation and discuss the effect of suboptimality.
\end{abstract}
\maketitle

\section{Introduction}
\label{sec:intro}

\begin{figure}[t!]
\includegraphics[width=8cm]{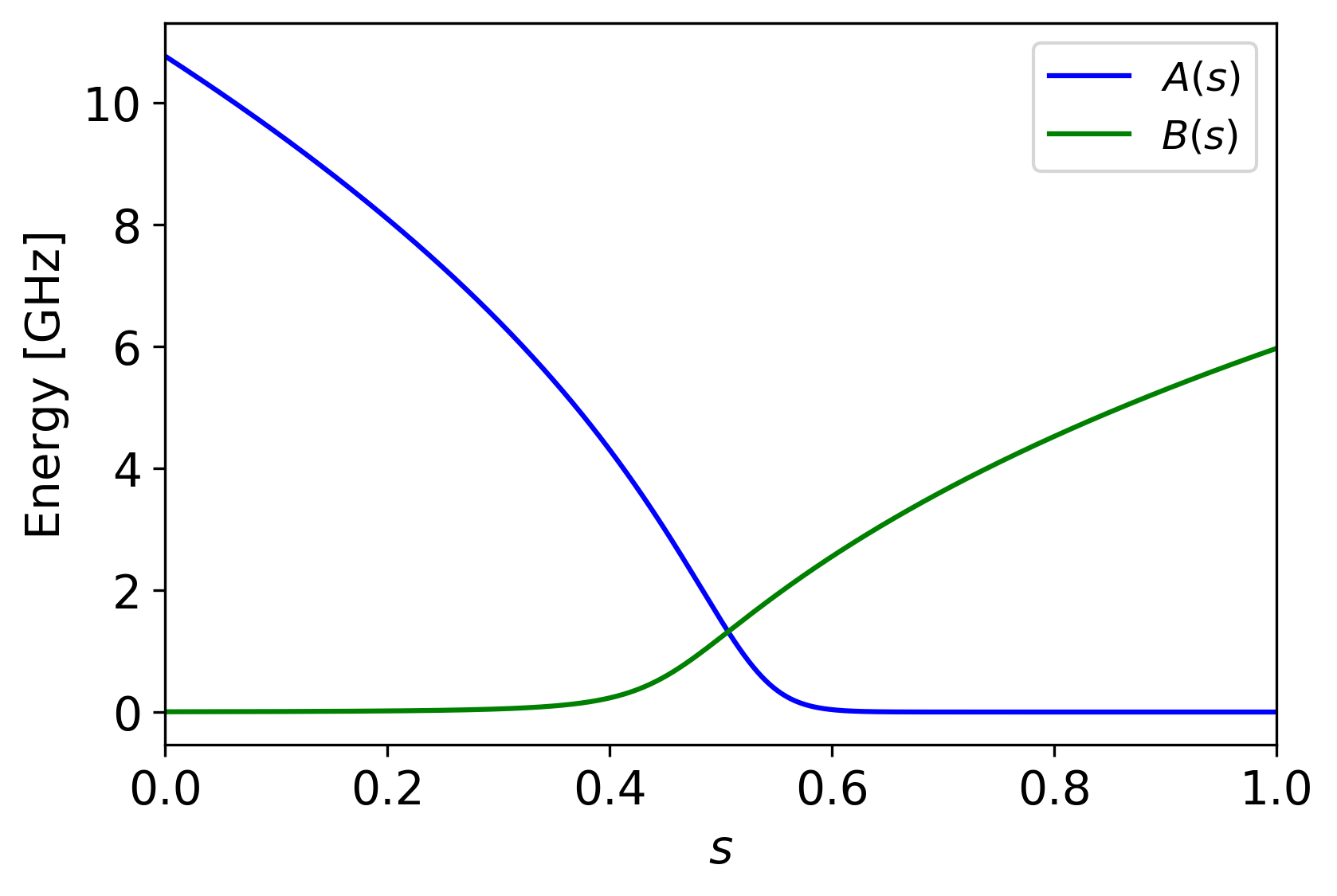}
\caption{\label{ABs} The envelope functions $A(s),B(s)$ as a function of the dimensionless annealing time $s$.}
\end{figure}

Quantum annealing (QA) \cite{1994CPL...219..343F, PhysRevE.58.5355, Brooke779} is a heuristic algorithm for finding low-energy configurations of Ising spin Hamiltonians, with applications in optimization and machine learning. Physical implementations of quantum annealers have matured to systems that include more than 5000 qubits, with increasing number of qubits expected in the future. The typical Hamiltonian implemented by these devices is
\begin{eqnarray}
    H(s) &=& A(s) H_D + B(s) H_P,  \label{H} \\
    H_D &=& -\frac{1}{2}\sum_i \sigma_i^{x}, \\
    H_P &=&  \sum_i h_i \sigma_i^z +\sum_{i<j} J^z_{ij} \sigma_i^z \sigma_j^z, \label{HP}
\end{eqnarray}
where $\sigma_i^{x,y,z}$ are the Pauli matrices acting on the $i$-th qubit, $H_D$ and $H_P$ are known as driver and problem Hamiltonians respectively, $h_i$ and $J_{ij}$ are dimensionless bias and coupling coefficients, and  $s = t/t_a$ is a dimensionless annealing time parameter with $t_a$ being the total annealing time. The envelope functions $A(s)$ and $B(s)$ are usually fixed by the experimental implementation; an example is plotted in Fig.~\ref{ABs}. 
Annealing is performed by initially letting the system relax to its ground state at $s=0$ when $A(s) \gg B(s)$, and then evolving to a configuration in which $A(s) \ll B(s)$ at $s=1$. Qubit states are measured at the end of annealing in the computation basis, which is defined by eigenfunctions of $\sigma^z_i$ denoted by $\ket{\uparrow}$ and $\ket{\downarrow}$ with eigenvalues $\pm 1$. 

For closed systems, the adiabatic theorem \cite{Kato1950,Amin2008,Lidar2009} ensures that the system remains in its ground state throughout the annealing if the evolution time is long relative to a time scale that is proportional to $1/\Delta^2$, where $\Delta$ is the minimum gap between the ground state and the first excited state \cite{Farhi2000}. Reading out the $N$ qubits then returns a configuration $\vec S \equiv \{S_1,S_2, ..., S_N \}$, with $S_i=\pm 1$, that minimizes the problem Hamiltonian $H_P$. In practice, the adiabatic theorem may be violated via fast evolution or thermal excitations, resulting in a suboptimal (but maybe still acceptable) solution. In this work, we only consider closed system evolution and take the ground state as the only acceptable solution.

The existing physical implementations of QA \cite{Johnson2011} use superconducting qubits coupled via only one degree of freedom (flux), giving rise to {\em stoquastic} Hamiltonians \cite{Bravyi2008,Bravyi2009,Marvian2019,2018arXiv180605405K}, i.e., Hamiltonians with no positive or complex off-diagonal elements. Equilibrium statistics of stoquastic Hamiltonians can be simulated with quantum Monte Carlo (QMC) methods with no \emph{sign problem} \cite{Loh1990,Troyer2005,Gupta2020}. QMC may also exhibit dynamical behavior similar to QA for special stoquastic Hamiltonians \cite{Isakov2016}, although this does not hold in general \cite{Hastings2013,Jarret2016,Andriyash2017}. Nonstoquastic Hamiltonians, however, are not treatable by QMC, hence their statistical and dynamical properties are extremely hard to simulate \cite{Marvian2019,Gupta2020}. They also can perform universal quantum computation \cite{Aharonov2008,Biamonte2008,Mizel2019} suggesting that nonstoquasticiy may be connected to quantum advantage in QA.

In order to make Hamiltonian \eqref{H} nonstoquastic, one needs to introduce interactions via other degrees of freedom, e.g., by changing the driver Hamiltonian to
\begin{equation}
    H_D = -\frac{1}{2}\sum_i \sigma_i^{x} + \sum_{\substack{i<j}} (J^x_{ij} \sigma_i^x \sigma_j^x + J^y_{ij} \sigma_i^y \sigma_j^y). \label{HDn}
\end{equation}
We refer to the last two terms as XX and YY interactions, respectively, in contrast to the ZZ interaction in $H_P$. For $J^y_{ij}=0$, the driver \eqref{HDn} has positive off-diagonal elements when $J^x_{ij}>0$. This is indeed the regime in which most studies has been done. It has been shown that this nonstoquastic Hamiltonian can significantly improve performance of QA, but only in special cases \cite{Hormozi2017,Nishimori2017,Susa2017,Albash2019,Durkin2019}. Moreover, a recent study \cite{Crosson2020} has found that the minimum gap in nonstoquastic QA generally increases by {\em de-signing} the Hamiltonian, i.e., making the Hamiltonian stoquastic by simply changing the sign of all positive off-diagonal elements. Therefore, whether or not nonstoquasticity can result in quantum annealing advantage over classical approaches remains an open question.

Less studied is nonstoquasticity due to YY-interaction. Recently a pair of superconducting flux qubits with nonstoquastic Hamiltonian was implemented by coupling them via both their flux and charge degrees of freedom  \cite{Ozfidan2019}. The resulting Hamiltonian had a driver of the form \eqref{HDn} that was dominated by YY-interaction. A nonzero $J^y_{ij}$ is special because it generates positive off-diagonal elements regardless of its sign and makes adiabatic path and performance variant under {\em spin reversal transformation} (SRT), defined in the next section. The goal of this paper is to systematically study the role of SRT in quantum annealing with nonstoquastic drivers.

\section{Spin reversal transformation}

Let us define a gauge transformation
\begin{eqnarray}
\ket{\psi} \to \ket{\psi'} = U \ket{\psi}, \qquad H \to H' = U H U^\dagger
\end{eqnarray}
with the unitary operator
\begin{equation}
U \equiv \prod_i (\sigma^x_i)^{(1-\alpha_i)/2}
\end{equation}
where  $ \vec \alpha \equiv \{\alpha_1,\alpha_2, ..., \alpha_N \}$ is a set of transformation parameters  with  $\alpha_i = \pm1$. The unitary operator $U$ flips the state of qubit $i$ if $\alpha_i=-1$, otherwise, it does nothing. The sign of each term in the Hamiltonian is adjusted so that the total energy remains unchanged. The transformed Hamiltonian  $H'$ has parameters
\begin{eqnarray}
     h'_i &=& \alpha_i h_i, \label{h'} \\
     J'^x_{ij} &=& J^x_{ij}, \label{Jx'} \\
     J'^y_{ij} &=& \alpha_i \alpha_j J^y_{ij} , \label{Jy'} \\
     J'^z_{ij} &=& \alpha_i \alpha_j J^z_{ij} .  \label{Jz'}
\end{eqnarray}
It has exactly the same spectrum and dynamical behavior as $H$, as expected for gauge transformations, and the returned solution is the transformation of the original solution:
\begin{equation}
S'_i = \alpha_i S_i \label{s'}.
\end{equation}
In practice, changing the sign of $J^y_i$ is nontrivial, at least for the physical implementation of Ref.~\cite{Ozfidan2019}. We define {\em spin reversal transformation} (SRT) as transformations \eqref{h'}-\eqref{s'} without \eqref{Jy'}, i.e., with $J'^y_{ij} = J^y_{ij}$. Therefore, SRT only transforms the classical part of the Hamiltonian.

For $J^y_{ij}=0$, SRT is a true gauge transformation and is not expected to affect the dynamics. Therefore, solving the problem with QA using $H$ or $H'$ should lead to exactly the same probability of success. However, if there exist systematic errors in parameter specifications of the physical Hamiltonian, the errors will not be transformed if we submit $H'$ instead of $H$ to the QA hardware. This means that SRT is not a true gauge transformation at the physical level and therefore is expected to affect the probability of success. In these situations, parameter specification error can be mitigated (averaged out to some extent) by running the problem with a set of SRTs, each parameterized by a randomly selected $\vec \alpha$.

When $J^y_{ij} \ne 0$, Hamiltonian \eqref{H} is not invariant under SRT, although the problem Hamiltonian still remains invariant (ignoring parameter misspecification). This means that while the classical problem being solved stays the same, the quantum path in the Hilbert space through which the solution is reached can heavily depend on the choice of SRT. Specially, the minimum gap can significantly change between transformations, resulting in a huge difference in success probabilities. Our goal is to find ways to select SRT intelligently so that the performance is improved.

The driver Hamiltonian \eqref{HDn} contains all tunneling terms in Hamiltonian \eqref{H}. Single qubit tunneling is through $\sigma^x_i$ operators, and the XX and YY terms contribute to two-qubit cotunneling events. The matrix elements of $\sigma^x_i\sigma^x_j$ and $\sigma^y_i\sigma^y_j$ between states with ferromagnetic (FM) and antiferromagnetic (AFM) orientations are given by
\ba
\langle \uparrow\downarrow | \sigma^x_i\sigma^x_j | \downarrow\uparrow\rangle = 1 &,& \quad 
\langle \uparrow\uparrow | \sigma^x_i\sigma^x_j | \downarrow\downarrow\rangle = 1 \\
\langle \uparrow\downarrow | \sigma^y_i\sigma^y_j | \downarrow\uparrow\rangle = 1  &,& \quad
\langle \uparrow\uparrow | \sigma^y_i\sigma^y_j | \downarrow\downarrow\rangle = -1
\ea
While $\sigma^x_i\sigma^x_j$ does not distinguish between FM and AFM orders, $\sigma^y_i\sigma^y_j$ has off-diagonal elements with opposite signs. To the lowest order perturbation in $A(s)/B(s) \ll 1$, the two-qubit tunneling amplitudes for FM and AFM correlations are given by 
\ba
\Delta_{\rm FM} &=& A(s) \left( {A(s) \over 4J^z B(s)} - J^x + J^y \right), \label{DFM} \\
\Delta_{\rm AFM} &=& A(s) \left( {A(s) \over 4J^z B(s)} - J^x - J^y \right). \label{DAFM}
\ea
The first term in each equation describes tunneling through two single-qubit tunneling processes via $\sigma^x$ operators. The last two terms, on the other hand, are contributions of direct two-qubit cotunneling via XX and YY interactions. Notice that with a negative $J^x$ (stoquastic), the XX coupling always increases the tunneling amplitude for both FM and AFM correlations. The YY interaction with $J^y>0$, however, increases (decreases) tunneling amplitude for FM (AFM) correlated qubits, due to constructive (destructive) interference. For a pair of coupled qubits with zero bias, Eqs~\eqref{DFM} and \eqref{DAFM} determine the size of the spectral gap between the ground and first excited states. Therefore, for the same magnitude of ZZ coupling, FM coupling has a larger spectral gap than AFM coupling when $J^y>0$, as experimentally demonstrated in Ref.~\cite{Ozfidan2019}. The same argument also holds for larger clusters of strongly coupled qubits; the spectral gap is largest when couplings are maximally FM. 

In problems with first order phase transition \cite{AminChoi}, the minimum gap is typically suppressed because the ground state jumps between two states that are separated by a large hamming distance. In these cases, a large cluster of qubits needs to flip between the two crossing local minima. The cluster's tunneling amplitude at the avoided crossing determines the size of the gap. This phenomenon was experimentally demonstrated in Ref.~\cite{Dickson2013}, using a crafted 16-qubit problem with extremely small gap. If qubits' transverse fields can be tuned individually, one can increase $\Delta$ by changing the adiabatic path either randomly \cite{FarhiRandomPath} or algorithmically \cite{Dickson2012}. The presence of XX and/or YY interactions allows for alternative ways of changing the adiabatic path. Especially with YY interaction, every SRT introduces a new path. Therefore, by randomly selecting SRTs, one can randomize the adiabatic paths until a path with large $\Delta$ is reached. One may also choose SRTs intelligently using algorithms similar to \cite{Dickson2012}, which use information about the states reached via previous taken paths. Here, however, we are interested in algorithms that do not need information from previous samples.


\section{Numerical simulations}

In this section, we explore the effect of SRT on performance of QA assuming realistic parameters. The Hamiltonian is taken to be \eqref{H} with experimentally motivated $A(s)$ and $B(s)$ plotted in Fig.~\ref{ABs}. All qubit couplings (XX, YY, or ZZ) are assumed to be according to the Chimera topology \cite{Chimera}. Based on the experimental observations of Ref.~\cite{Ozfidan2019}, we choose $J^y_{ij} = 0.5$ whenever they are nonzero. Also, to allow a direct comparison between stoquastic and nonstoquastic drivers, we choose $J^x_{ij} = -0.5$ (if nonzero). We calculate the minimum gap during the evolution using exact diagonalization. We also calculate the probability of success by solving Schr\"odinger equation with annealing time $t_a = 1\,\mu$s.

\subsection{Crafted problem}

\begin{figure}[t!]
\includegraphics[width=8cm]{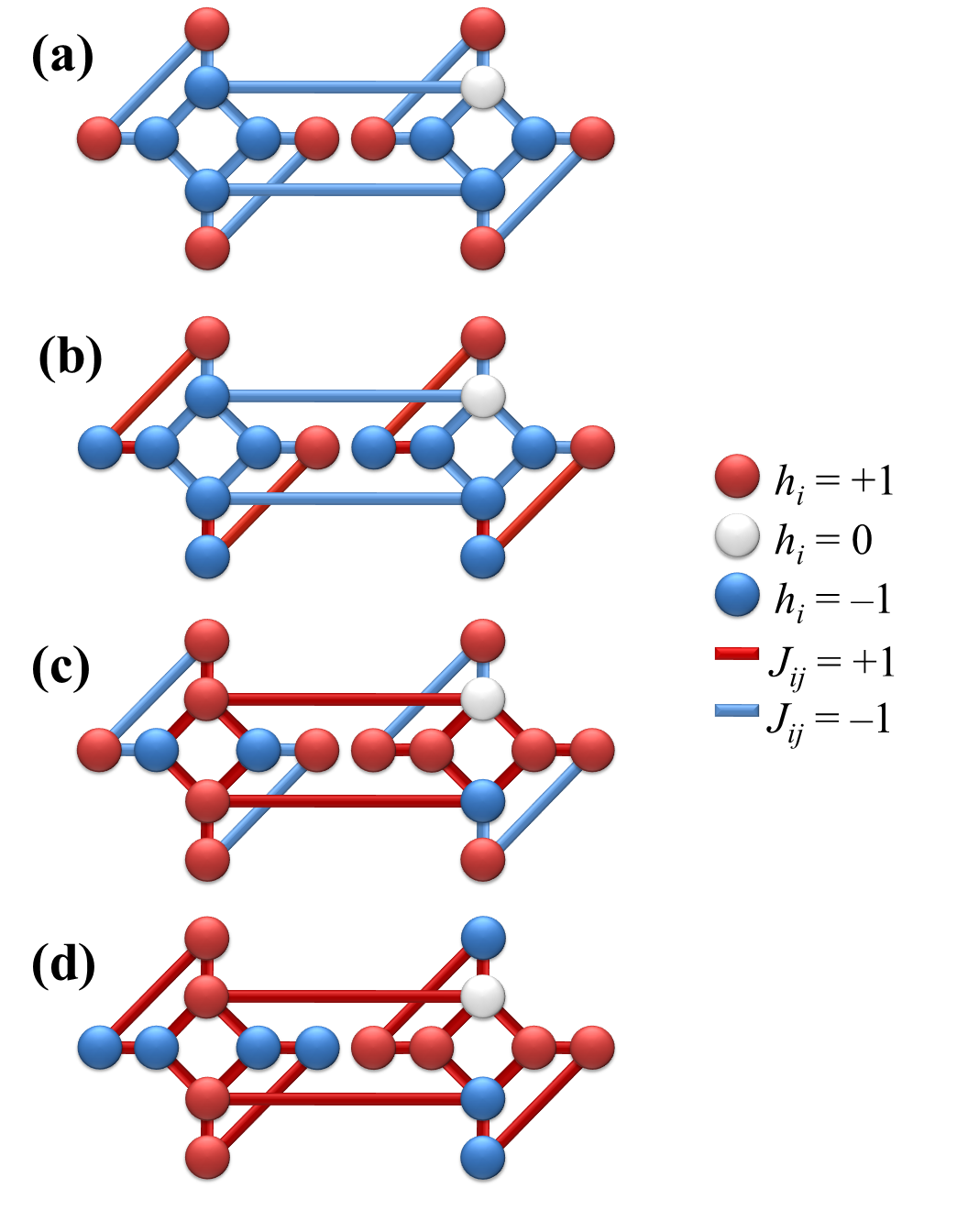}
\caption{\label{Dickson} (a) A crafted problem designed to have a small minimum gap. All couplers are ferromagnetic. (b)-(d) Three spin reversal transformations of problem (a) with some or all couplers antiferromagnetic.}
\end{figure}

\begin{figure}[t!]
\includegraphics[width=9cm]{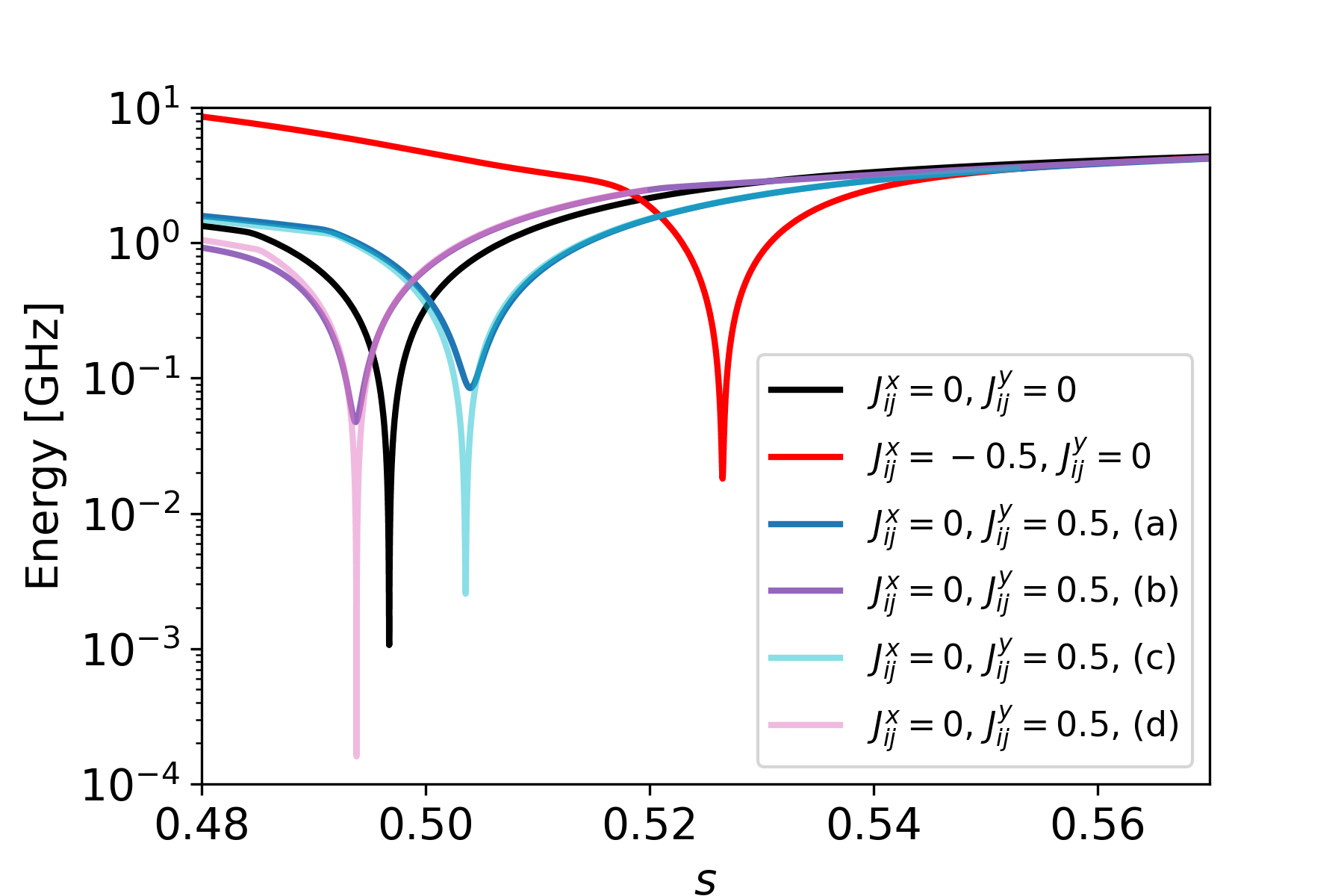}
\caption{The energy splitting between ground and  first excited state as a function of the annealing parameter $s$ for the problems in Fig.~\ref{Dickson} with different driver Hamiltonians. }
\label{Spectrum} 
\end{figure}

We first consider a slightly modified version of the problem studied in \cite{Dickson2013}, as shown in Fig.~\ref{Dickson}(a). Parameters of the problem Hamiltonian are color coded in the figure. We divide the qubits into two groups, inner qubits (blue and white circles) and outer qubits (red circles). All couplings are FM, therefore, the two ferromagnetically oriented states $\ket{\uparrow\uparrow ... \uparrow}$ and $\ket{\downarrow\downarrow ... \downarrow}$ are energetically favored by the coupling terms. The biases in Fig.~\ref{Dickson}(a) are four positive, three negative, and one zero, making  $\ket{\downarrow\downarrow ... \downarrow}$ the unique ground state and $\ket{\uparrow\uparrow ... \uparrow}$ an excited state. The outer qubits are pairwise coupled. Each outer qubit is in agreement with its applied bias in the ground state, but opposes the bias in the above excited state. If in the excited state a pair of coupled outer qubits are flipped together, two bias terms will be satisfied but two couplers will be violated, leaving the energy unchanged. Therefore, with the existing 4 outer pairs there are $2^4 = 16$ degenerate excited states all connected by two-qubit flips. This degeneracy is lifted by the transverse field. Each coupled pair would lower their energy by forming an entangled state $(\ket{\uparrow\uparrow} + \ket{\downarrow\downarrow})/\sqrt{2}$. The lowest excited state is therefore a superposition of these 16 degenerate states. As transverse field is increased (moving back in $s$), the splitting of the excited states grows until the lowest excited state crosses the ground state. The minimum gap at this avoided crossing is proportional to the tunneling amplitude between the two (localized) crossing states. Each of the 16 classical states in the superposition has a large hamming distance to the ground state (8 to 16 bit flips), resulting in a very small $\Delta$. Figure \ref{Spectrum} shows the energy splitting between the ground and first excited states for Hamiltonian  (\ref{HDn}) during the annealing according to the schedule in Fig.~\ref{ABs}. As expected, the minimum gap for the original problem, with no YY or XX interactions (black curve), is very small, $\Delta_0 \approx 10^{-3}$ GHz (the index $0$ indicates $J_{ij}^x=J_{ij}^y = 0$ for every i,j). The gap is significantly increased when XX-interaction is turned on (red curve in Fig.~\ref{Spectrum}).

In the presence of YY-interaction, $\Delta$ is expected to depend on SRT. Figures \ref{Dickson}(b)-(d) show three SRTs of Fig.~\ref{Dickson}(a). The eight inner qubits all flip at the avoided crossing, therefore, they need to be coupled ferromagnetically to allow maximum tunneling amplitude. These couplings are turned from FM to AFM in Fig.~\ref{Dickson}(c). As it is clear from Fig.~\ref{Spectrum}, the size of the minimum gap for problem (c) is reduced by almost two orders of magnitude compared to (a) although its position remains almost unchanged.  The outer qubits, on the other hand, determine the position of the avoided crossing. Their pairwise tunneling is what lifts the degeneracy of the 16 classical excited states and creates the avoided crossing. Therefore, turning the coupling between the outer qubits from FM to AFM should reduce the splitting of the degenerate states and push the avoided crossing back toward a smaller $s$, as is the case in Fig.~\ref{Spectrum} for (b) and (d) curves. Since the transverse field is larger earlier in the anneal, one might expect the minimum gap to be larger for problem (b) compared to (a), and likewise (d) compared to (c). However, Fig.~\ref{Spectrum} shows the opposite behavior. This is because not only the inner qubits, but also some of the outer qubits flip between the ground state and each of the 16 degenerate excited states. At a fixed transverse field, the multi-qubit tunneling amplitude is largest when the outer qubits are coupled ferromagnetically. Since the transverse field is not fixed, the two effects compete with each other; coupling the outer qubits antiferromagnetically pushes the avoided crossing to a smaller $s$ hence increasing in the transverse field, but the increase is not enough to compensate the reduction of multi-qubit tunneling amplitude due to AFM coupling. As a result, the largest minimum gap happens when all couplers are FM, as they are in problem (a).

\begin{figure}[t!]
\includegraphics[width=9cm]{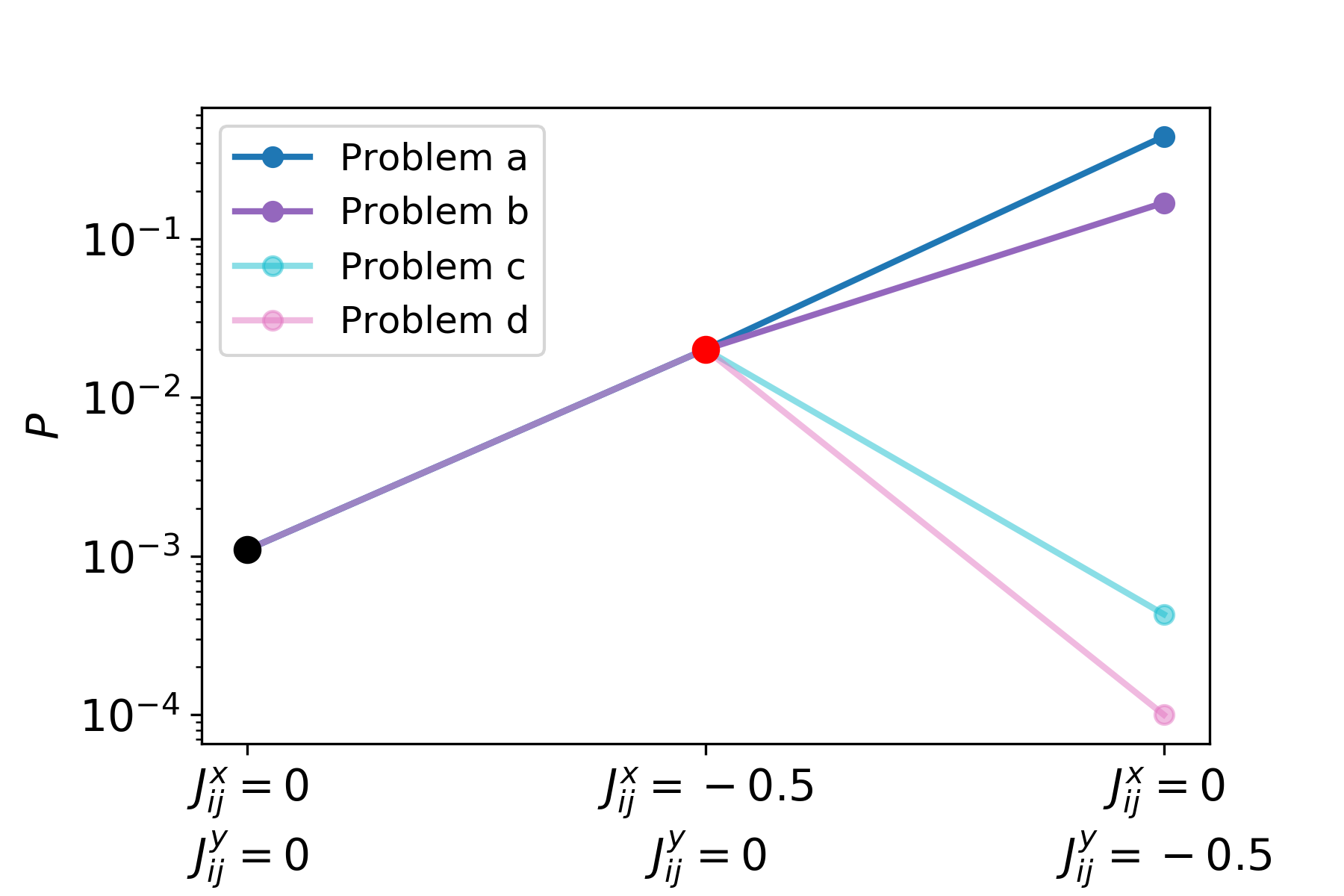}
\caption{\label{Fig:Min_gaps} The success probabilities for the problems in Fig.~\ref{Dickson} with different driver Hamiltonians. }
\end{figure}

Figure \ref{Fig:Min_gaps} compares probabilities of success for the scenarios presented in Fig.~\ref{Spectrum}. As expected, the success probabilities in Fig.~\ref{Fig:Min_gaps} are small and correlate with the minimum gap sizes in Fig.~\ref{Spectrum}. While the XX-coupling enhances the performance regardless of the SRT, the YY-coupling may increase or decrease the probability of success depending on the SRT. The best performance is obtained for problem (a) when YY-interaction is on. Since the magnitudes of $J^x_{ij}$ and $J^y_{ij}$ are the same, switching from YY to XX interaction is equivalent to one of the de-signing processes proposed in \cite{Crosson2020}. Clearly from problem  (a) to (b) de-signing did not improve the performance, in contrast to Ref.~\cite{Crosson2020}. Therefore, although nonstoquasticity does not automatically lead to advantage in QA, with a right choice of SRT a nonstoquastic Hamiltonian with YY interaction can outperform the original  stoquastic Hamiltonian as well as the de-signed stoquastic Hamiltonian (with $J^x_{ij} = -J^y_{ij}$).

\subsection{Random problems}

We now investigate whether the observations in the previous example hold for random problems. We generate problems by randomly selecting $J^z_{ij}$ from $[-1, 1]$ with 4 bits of precision (16 evenly distributed values within that range). The biases are also selected in the same way or all taken to be zero. The connectivity graph is again Chimera, but with $N=12$ (six qubits within each unit cell) to limit the computation time. Since most generated problems of this size are easy, we only keep the ones with small gap: $\Delta < 0.1$. In total we generated 100 problems with random $h_{ij}$ and 30 problems with $h_{ij} = 0$. Since we did not find any qualitative difference between the two cases, we combine them into a single set and present them together.

The best performance in the previous crafted example was obtained when the couplers were all FM. In random problems, however, the coupling terms are usually frustrated, meaning that no solution can satisfy them all simultaneously. Therefore, no SRT can make all the couplers FM, i.e., making a frustrated problem unfrustrated. Since qubits with strongest couplings are most likely to be correlated, it is reasonable to choose SRTs that make those couplers FM and allow the weak ones to be AFM. We define the average coupling strength for the transformed Hamiltonian as
\be
\bar J = {1 \over N_J} \sum_{i,j} J'^z_{ij},
\label{Jbar}
\ee
where $N_J$ is the number of couplers. Clearly, $\bar J$ varies with SRT and the more negative it is the more ferromagnetic the couplings are.

\begin{figure}[t!]
\includegraphics[width=9cm]{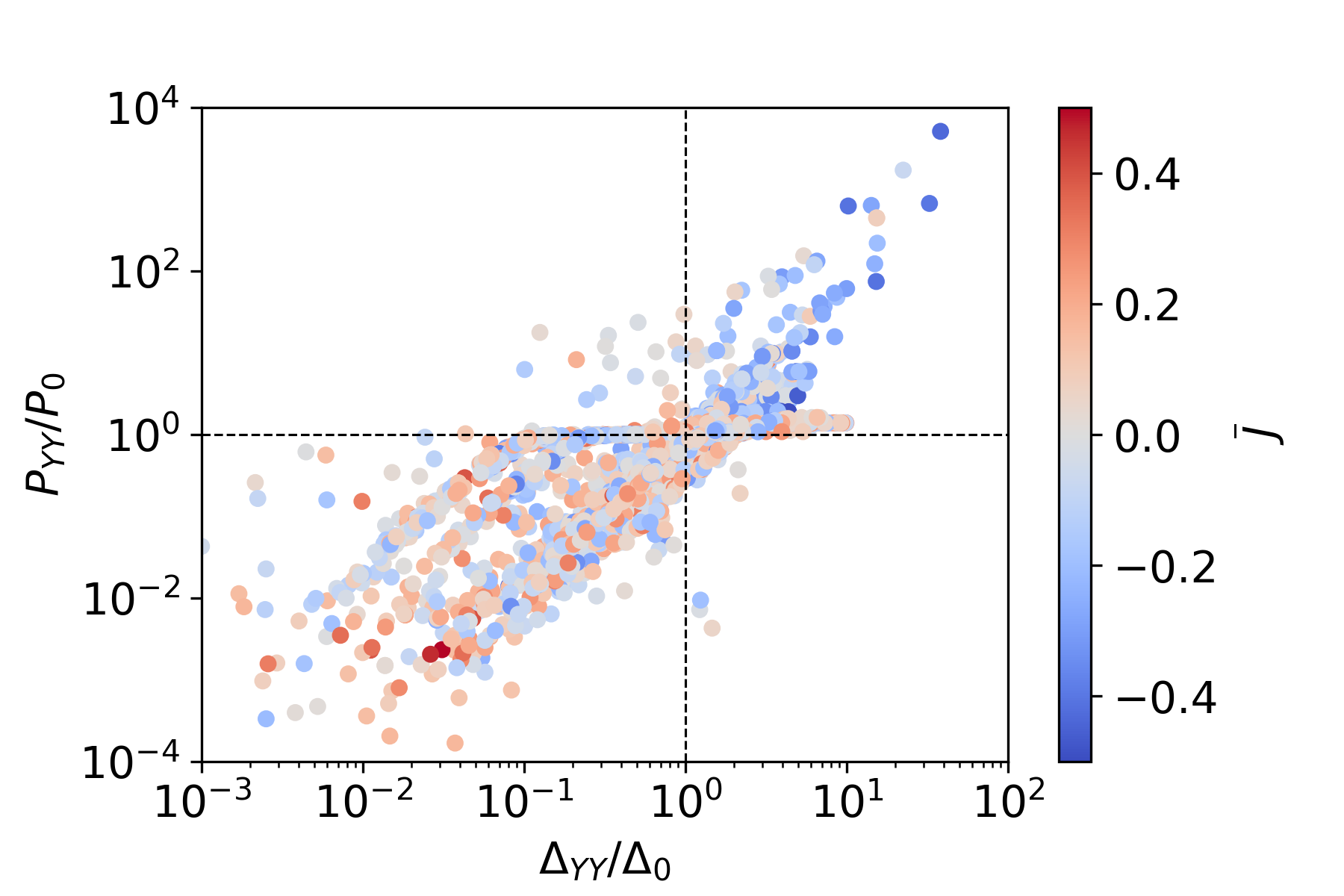}
\caption{\label{Fig:Pyy_vs_P0_random} Scatter plot of the  relative success probability and the relative minimum gap size corresponding to the driver Hamiltonian \eqref{HDn} with $J_{ij}^x = 0$ and $J^y_{ij} = 0.5$. For each of the 130 random problems we have applied 50 random SRTs. The color-code shows the value of $\bar{J}$ for each instance. Majority of cases with improved performance have $\bar{J}<0$. }
\end{figure}

For each of the 130 generated random problems we choose 50 SRTs by randomly selecting $\alpha_i$ from $\pm 1$ with equal probability. Figure \ref{Fig:Pyy_vs_P0_random} shows a scattered plot of success probability versus minimum gap in the presence of YY-interaction for all the 65,000 instances. The probabilities and the gap values are normalized to their corresponding values in the original problem. Therefore, a value bigger than 1 means improvement. A correlation between the probabilities and the minimum gap sizes can be recognized in the Figure. The color-coding represents the average coupling strength $\bar J$. As in the previous example, adding YY-interactions can improve or impair the performance depending on the SRT. However, as the color coding indicates, there is a close correlation between the performance and the sign of $\bar J$; most problems with FM dominated couplers ($\bar J<0$) show improvement. 

\begin{figure}[t!]
\includegraphics[width=9cm]{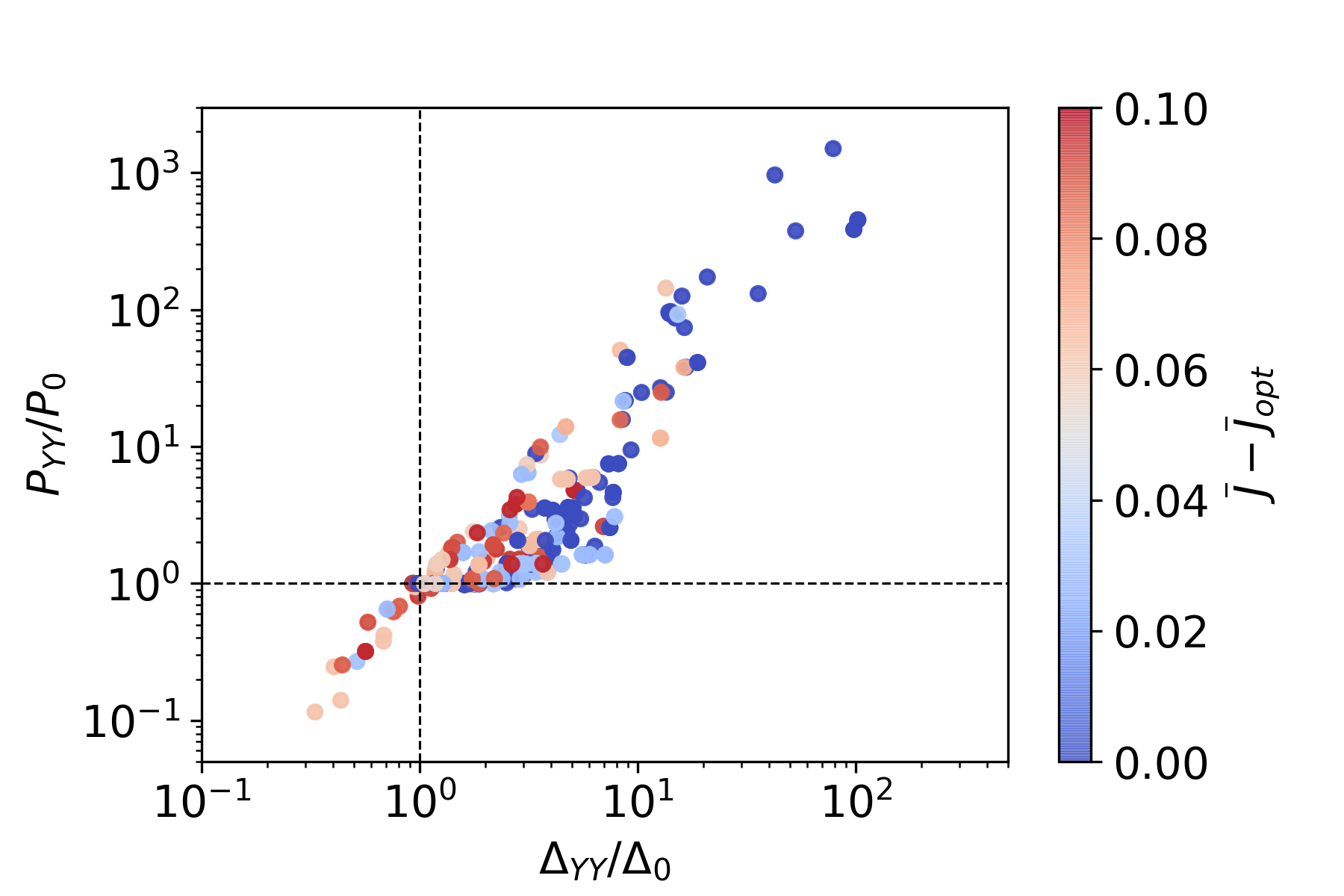}
\caption{
Scatter plot of relative success probabilities and relative minimum gaps for SRTs obtained by minimizing Eq.~\eqref{Ealpha}. Color coding represents the distance from the optimum. We have only kept suboptimal solutions with $\bar J - \bar J_{\rm opt} < 0.1$.
\label{Fig:Prel_vs_Deltarel_comp} }
\end{figure}

The above observation as well as the results of the previous example suggest that a SRT that minimizes $\bar J$ is likely to improve the performance. Using Eq.~\eqref{Jz'} and \eqref{Jbar}, we write the objective function as
\be
\bar{J} (\vec \alpha) = {1 \over N_J} \sum_{i,j} J^z_{ij} \alpha_i \alpha_j.
\label{Ealpha}
\ee
The solution $\vec \alpha_{\rm opt}$ that minimizes \eqref{Ealpha} defines the desired SRT. Since the number of variables is not very large ($N=12$), we can find all global and local minima of \eqref{Ealpha} through exhaustive search. This, however, is not possible for larger problems. Since $\alpha_i$ is a binary variable with values $\pm 1$ (similar to $S_i$), the objective function \eqref{Ealpha} itself is an Ising problem. Indeed, \eqref{Ealpha} forms the quadratic part of the problem Hamiltonian \eqref{HP} and is equivalent to $H_P$ when $h_i=0$. Finding $\vec \alpha_{\rm opt}$ is therefore NP-hard, as complex as minimizing the original problem Hamiltonian. However, the quantum annealer itself can be used to minimize Eq.~\eqref{Ealpha}. Especially for problems with $h_i = 0$, the optimal solution to the original problem $H_P$ and the optimal SRT coincide: $\vec S_{\rm opt} = \vec \alpha_{\rm opt}$, Therefore, solving the problem itself with QA gives the SRT for the next run and the process can be repeated iteratively until the desired solution is reached. Moreover, suboptimal solutions to \eqref{Ealpha} also improve the performance and could be as good as, and sometimes even better than, the optimal solution, as we show below.

\begin{figure}[t!]
\includegraphics[width=9cm]{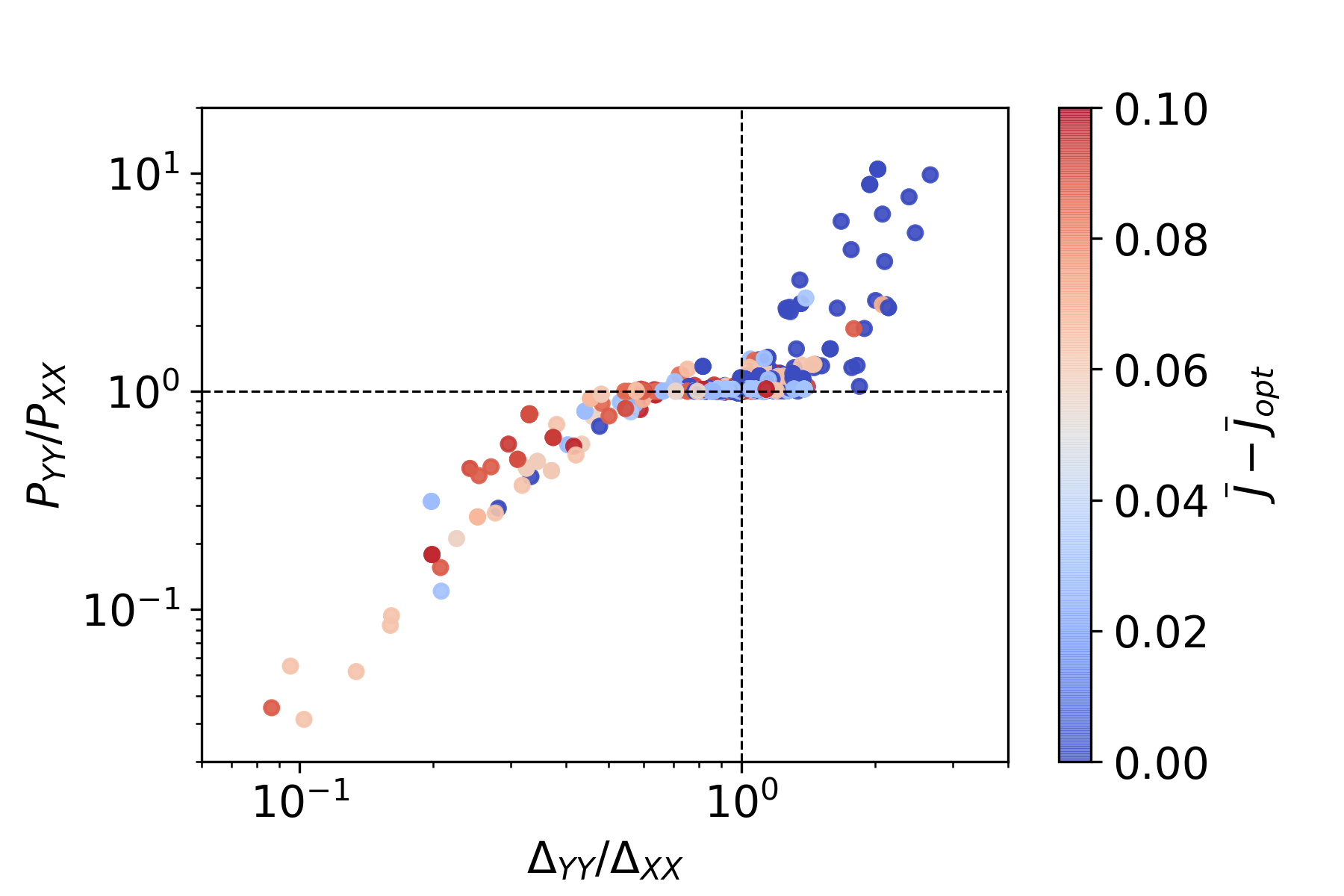}
\caption{
Comparison between nonstoquastic Hamiltonian with YY-interaction ($J_{ij}^x = 0$ and $J^y_{ij} = 0.5$) and the (de-signed) stoquastic Hamiltonian obtained by replacing YY with XX-interaction ($J_{ij}^x = -0.5$ and $J^y_{ij} = 0$). The instances are the same as in Fig.~\ref{Fig:Prel_vs_Deltarel_comp} with colors representing the distance from the optimum. 
\label{Fig:Prel_vs_Deltarel_comp_XX} }
\end{figure}

Figure \ref{Fig:Prel_vs_Deltarel_comp} shows a scatter plot similar to Fig.~\ref{Fig:Pyy_vs_P0_random}, but instead of random, the SRTs are obtained via optimization of Eq.~\eqref{Ealpha}. Colors represent distance from the optimal $\bar J_{\rm opt}$. We only kept local minima with $\bar{J} - \bar{J}_{opt} < 0.1$. As it is evident from the figure, for majority of the 130 problems, the minimum gap size is increased for the SRT corresponding to the optimal or suboptimal solutions of Eq.~\eqref{Ealpha}, and the probability of success is improved by up to more than three orders of magnitude. There were also cases that suboptimal solutions to  Eq.~\eqref{Ealpha} gave better probability of success than the optimal one. Table I shows the percentage of improved cases: $91\%$ of the optimal solutions and $83\%$ of the suboptimal solutions led to SRTs that increased the success probability.

\begin{table}[]
\caption{Percentages of improved instances for suboptimal and optimal solutions of Eq. (\ref{Ealpha}).}

\begin{tabular}{l|ll}
\hline
$\bar{J} - \bar{J}_{opt}$ & $P_{YY}/P_0$ & $P_{YY}/P_{XX}$  \\
\hline \hline
 Suboptimal  & 83\%                   & 56\%  \\
Optimal  & 91\%                    & 82\%   \\
\hline
\end{tabular}
\end{table}

Finally, we investigate how a nonstoquastic Hamiltonian with YY-interaction compares, in terms of QA performance, to the corresponding stoquastic (de-signed) Hamiltonian with XX-interaction. The magnitude of the YY and XX interactions are the same in the two Hamiltonians ($|J^x_{ij}| = |J^y_{ij}| = 0.5$). Since for YY-interaction the adiabatic path depends on SRT, we expect the performance to be better or worse than the stoquastic case depending on the SRT. Indeed, we find that for majority of random SRTs, stoquastic Hamiltonians yield better performance than the corresponding nonstoquastic ones, in agreement with Ref.~\cite{Crosson2020}. However, when we find optimal SRTs by minimizing Eq.~\eqref{Ealpha},the nonstoquastic Hamiltonians on average outperform the corresponding stoquastic ones. Figure \ref{Fig:Prel_vs_Deltarel_comp_XX} plots the relative probabilities and minimum gap sizes for the 130 random problems studied before.  The percentages of improvement are also reported in Table I. Quantum annealing with stoquastic Hamiltonian (XX-interaction) is outperformed by the corresponding nonstoquastic one (YY-interaction) for 82\% of the problems when the optimal SRT (the global minimum of Eq.~\eqref{Ealpha}) was applied and 56\% of cases when suboptimal SRTs were applied.

\section{Conclusions}

We investigated the effect of nonstoquastic Hamiltonian with YY-interaction between the qubits on performance of quantum annealing. The existence of YY-interaction makes the adiabatic path and the performance strongly dependent on SRT. Random transformations in general do not improve the performance. We found that the transformation that makes the average ZZ-coupling maximally ferromagnetic is most likely to improve the performance, sometimes by several orders of magnitude. The improvement is expected to be even larger at larger sizes; numerical simulations become intractable due to nonstoquasticity. Suboptimal solutions also improve the performance, but with less frequency. We should mention that the SRT obtained by minimizing the average coupling is not the best possible SRT among all the exponentially large number of possible transformations. It is conceivable that more elegant algorithms, e.g., that use information from previously obtained solutions, lead to better transformations. One may also use machine learning techniques to choose SRTs based on the numerical observations or experimental data once large scale quantum annealers with nonstoquastic interactions are built. 

In the physical implementation of Ref.~\cite{Ozfidan2019}, nonstoquasticity was obtained by adding capacitive coupling to magnetically coupled flux qubits. The resulting Hamiltonian had an extra XX-coupling in addition to the expected YY-coupling. The XX-coupling was stoquastic with a magnitude that depended on ZZ-interaction. Both XX and YY couplings favored FM over AFM correlation in terms of contribution to two-qubit tunneling. As a result, the dependence on SRT is expected to be stronger than that for YY-coupling alone. Moreover, the addition of coupling capacitors will increase the total capacitance of each qubit, resulting in a smaller tunneling amplitudes. However, only a few percent reduction is expected, not large enough to eliminate the orders of magnitude enhancement of performance observed above. More research is needed to assess the value of such capacitive interactions in practical quantum annealers at large scales.

\section*{Acknowledgement}
The authors are thankful to E. Andriyash, H. Nishmori, J. Raymond, and A. Smirnov for fruitful discussions. EML acknowledges support from her NSERC-CREATE IsoSiM Fellowship and AZ was partly supported by the Mitacs Accelerate program.





\end{document}